\documentclass[prc,aps,nofootinbib,showkeys,showpacs,twocolumn]{revtex4}
\usepackage{epsfig}
\usepackage{graphicx}
\usepackage{amssymb}
\usepackage{mathtools}
\usepackage{color}
\usepackage{bm}
\usepackage{amsmath}
\usepackage[colorlinks=true]{hyperref}
\usepackage{cleveref}
\crefname{equation}{}{}

\newcommand{\method}{MC-TDHFB\ }
\newcommand{\methodi}{MC-TDHFB$_{\textrm{I}} $\ }

\graphicspath{{figures/}}

\begin{document}

\title{Microscopic description of pair transfer between two superfluid systems (II):
a quantum mixing of Time-Dependent Hartree-Fock Bogolyubov trajectories
}

\author{David Regnier} \email{regnier@ipno.in2p3.fr}
\affiliation{Institut de Physique Nucl{\'e}aire, IN2P3-CNRS, Universit{\'e} Paris-Sud,Universit{\'e} 
Paris-Saclay, F-91406 Orsay Cedex, France}

\author{Denis Lacroix} \email{lacroix@ipno.in2p3.fr}
\affiliation{Institut de Physique Nucl{\'e}aire, IN2P3-CNRS, Universit{\'e} Paris-Sud,Universit{\'e} 
Paris-Saclay, F-91406 Orsay Cedex, France}
\date{\today}
\begin{abstract}
While superfluidity is accurately grasped with a state that explicitly breaks the particle number symmetry, a precise description of phenomena like the particle transfer during heavy-ion reactions can only be achieved by considering systems with good particle numbers.
We investigate the possibility to restore particle number in many-body dynamical problems by mixing-up several Time-Dependent 
Hartree-Fock Bogolyubov (TDHFB) trajectories. In our approach, each trajectory is independent from the others and  the quantum mixing between trajectories is deduced from a variational principle. The associated theory can be seen as a simplified version of the Multi-Configuration TDHFB (MC-TDHFB)  theory.  Its accuracy to tackle the problem of symmetry restoration in dynamical problems is illustrated for the case of 
two superfluid systems that exchange particles during a short time. In Ref. [Phys. Rev. C 97, 034627 (2018)], using a schematic model where two systems initially described by a pairing Hamiltonians are coupled during a short contact time, it was demonstrated that statistical mixing of TDHFB trajectories can only qualitatively describe the transfer process and that a fully quantum treatment is mandatory. 
We show here that the present MC-TDHFB approach gives an excellent agreement with the exact solution when the two superfluids are the same (symmetric case) or different (asymmetric case) and from weak to strong interaction strength.
Finally, we discuss the benefits and bottleneck of this method in view of its application to realistic systems.
\end{abstract}



\maketitle

\section{Introduction}

Symmetry breaking states are widely used in nuclear physics to describe the complex correlations in atomic nuclei at a low numerical cost. 
A widespread and successful approach in describing the low energy properties of  atomic nuclei is the nuclear energy density functional (EDF)~\cite{Ben03,Rin81}. In its simplest form, the system is represented by one symmetry breaking 
Slater determinant or more generally quasi-particle vacua that minimizes a nuclear energy density functional.
The most commonly broken symmetries are the translational invariance, the rotational symmetry for deformed systems and the $U(1)$ symmetry associated with the number of particles.
Allowing for such symmetry breaking captures efficiently the gross features of nuclei (e.g. its binding energy, radius, deformation,...) with rather simple density functionals.
However, quantitative prediction of the low--lying spectroscopy of atomic nuclei requires a theory that respects the symmetry of the nuclear Hamiltonian.
In that direction, important efforts are devoted in mixing several quasi-particle vacua into a Multi-Configuration (MC) framework, also called Generator Coordinate Method (GCM). This approach allows for instance for  
the prediction of the spin/parity of low--lying excited states as well as the inter-level transition rates~\cite{Egi16,Ben08}.
A generalization of this strategy to the time--dependent problem is highly desirable to describe nuclear reactions.
Up to now, developments in this direction were heavily hindered by the numerical cost of treating explicitly the time dimension.
In the last few years, several groups managed to describe for the first time heavy--ion collisions and fission within time-dependent EDF 
(TD-EDF) calculations that (i) includes the nucleon-nucleon pairing residual interaction, (ii) do not assume any built-in symmetry for the computation of single particle wave--functions~\cite{Sca15,Bul16,Mag17}.
As the computational power increases and numerical methods improve, the development of multi-configuration framework for reactions
seems timely.

One of the topics of current interest that would benefit from such a machinery is the question of  the role played by the nuclear superfluidity in the transfer and fusion channels of heavy--ion collisions at sub--Coulomb barrier energies. 
Recently, a number of experimental investigations have been 
made to study the competition between deep sub-barrier fusion and transfer channels~\cite{Cor11,Mon14,Mon16,Raf16}. These experimental investigations 
are made in parallel with numerous theoretical works (see for instance the recent review of Ref. \cite{Sek19}). 
Such a precise knowledge would also enable us to constrain the pairing interactions used in EDF approaches.

Recent attempts to answer this question (two colliding superfluid nuclei) in the framework of TDHFB were lead by Hashimoto \textit{et al.}~\cite{Has16} and Magierski \textit{et al.}~\cite{Mag17}. In this theoretical framework, it was shown that the properties of collisions between two superfluids atomic nuclei depend on the phases of the pairing field at play in each nucleus at initial time.  
In particular, varying the initial relative phase between the two systems induces variations of the fusion barrier energy height and impacts the number of particles exchanged during the reaction. The magnitude of this effect associated to superfluidity may lead to several 10 MeV change in the fusion barrier height for collisions between two $^{90}$Zr nuclei~\cite{Mag17} (see also recent precise discussion in \cite{Sca18}).
These variations might however be spurious consequences of the particle number symmetry breaking inherent to the HFB and/or TDHFB framework. 
This shortcoming can also be seen as TDHFB making a semi-classical treatment only of the gauge angle which plays here the role of a collective variable.

A proper treatment of collective degrees of freedom requires a priori to re-quantize them, leading automatically to a Multi-Configuration
framework with quantum mixing of several many-body states. Such a mixing however significantly increases the numerical effort and most 
often is at the limit or beyond our actual computational  capacities.  An intermediate semi-classical approximation consists in assuming 
that the quantum fluctuations can be treated simply as a statistical mixing. This technique is generically called phase-space approach hereafter.
It appears to be quite natural when making contact with experiments. For instance, when considering collisions 
between deformed nuclei, a statistical treatment corresponds to perform average over the angle between the deformation axis 
and the beam axis \cite{Ayi10}. The situation is similar for particle number where 
at least a statistical average of different orientations in gauge space should be performed. 
Another illustration 
of the statistical treatment is the stochastic mean-field theory 
\cite{Ayi08,Lac14}.  This approach has been shown to be particularly successful  to describe the dynamics close to a spontaneous symmetry breaking \cite{Lac12} and can eventually describe the physics of superfluid systems \cite{Lac13}. 
However, by construction, statistical hypothesis neglects possible quantum interferences  between mean-field trajectories. 
Coming back to describing collisions between two superfluids, the quantitative estimates of superfluity effects on reactions observables 
was made using a semi-classical phase-space average method in \cite{Mag17}. The validity of phase-space approaches for the problem of two superfluid nuclei transferring particles has been questioned in
Ref. \cite{Reg18}. We have shown that the brute force phase-space method based on statistical average of TDHFB trajectories 
can qualitatively reproduce some aspects of transfer but fails to give a precise description of the particle transfer process (see also discussion in Ref. \cite{Bro91}).

The aim of the present study is to describe the same problem including quantum mixing of different TDHFB trajectories.   
The proper treatment of quantum effects in collective space requires to re-quantize the mean-field evolution, a problem that is know to be of extreme complexity \cite{Bal69,Rei80,Rei83,Rei87,Dro86,Zha90,Nak16,Fan18-a,Fan18-b}. 
The strategy we use here is to start from a variational principle where the trial state is written in terms of a quantum mixing of several 
evolving HFB vacua. 
This strategy is similar to the Time-Dependent Generator Coordinate Method (TDGCM) that is used for instance to describe the fission process~\cite{Reg16,Tao17,Zde17}. 
It should however be noted that in state of the art TDGCM applications, states are assumed time-independent. In addition, they are mostly chosen on the adiabatic energy landscape except for a few attempts to include the collective impulsions~\cite{Goe80} or some intrinsic energy in the form of quasi-particles excitations~\cite{Ber11} or temperature~\cite{Zha19}.
This often restricts the approach to describe slow collective process. We consider here the possibility that the states evolve in time.   
In the following, to make contact with other fields \cite{Mey90,Mey09,Hoc11}, we will generically 
call such configuration-mixing with evolving states,  Multi-Configuration TDHFB. Note that, as far as we know, other fields of physics make use  of a similar approach but without the $U(1)$ symmetry breaking, leading to the so-called MC-TDHF. 
G. Scamps \textit{et al.} attempted in Ref.~\cite{Sca17,Sca17b} to develop a simplified theory where several TDHFB evolutions are mixed in time. While appealing, the proposed approach suffers from the lack of prescription to determine the relative phase between evolving vacua leading to uncontrolled fluctuations in the observables.  In the present work, based on the pioneering work of P.-G. Reinhard et al \cite{Rei83}, we show that a MC-TDHFB approach where the wave-function is represented by a quantum mixture of independent TDHFB trajectory can be accurately used to tackle the problem of symmetry restoration in dynamical problem and predict accurately the transfer of particles between two superfluid systems.  

In Sec.~\ref{sec:general}, basic ingredients of the MC-TDHFB methods are presented. Sec.~\ref{sec:application} is devoted to the application of MC-TDHFB to the model case. Comparison is made with the exact solution and the phase-space combinatorial approach 
developed in our previous work \cite{Reg18}.

\section{Discussion on the \method method}
\label{sec:general}

\subsection{General formulation of the \method framework}

We consider the generic ansatz where the system is described at all time $t$ as a quantum superposition of several time-dependent quasi-particle vacua $\{ \phi_\alpha (t)\}$:
\begin{eqnarray}
 |\psi(t)\rangle =\sum_\alpha f_\alpha(t)  |\phi_\alpha (t)\rangle, \label{eq:mix}
\end{eqnarray}  
with mixing coefficients  $\{ f_\alpha(t) \}$ that also depend on time. 
A natural way to determine the evolution is to use the Dirac-Frenkel variational principle that would impose the action
\begin{align}
 S = \int_{t=t_0}^{t_1} \langle \psi(t) | H - i\hbar \partial_t | \psi(t)\rangle dt
 \label{eq:varia}
\end{align}
to be stationary. 
The most general (and complex) situation consists in making the variation both with respect to the mixing coefficients and with respect to the 
quasi-particle components forming the states $|\phi_\alpha(t) \rangle$.
Noting that quasi-particle vacua form an over-complete basis of the many-body problem \cite{Bla86}, it might seems surprising 
to use Eq.~(\ref{eq:mix}) with time-dependent states since a time-independent basis can be used to solve the problem exactly. 
The real usefulness of having time-dependent states instead of time-independent 
stems from the possibility to describe certain physical effects with a reduced number of many-body states compared to the case where time-independent states are used
\footnote{
A seminal example is the description of giant resonance in nuclei. In the TDHF (or TDHFB) approach, a single 
time-dependent state provides a rather accurate description of the collective response. The same problem can be solved using coupled channels equations on a fixed multi-particle multi-hole basis built on the HF (or HFB) ground state. However, the numerical effort as well as the  
size of the required many-body Hilbert space would be much larger.   
}.

This general strategy has been declined into several flavors within the different fields of physics. In quantum chemistry, where the $U(1)$ symmetry is not broken, the quasi-particle vacua for the electrons reduce to Slater determinants and are usually built as different configurations (occupation numbers) on the same set of orthonormal orbitals (single particle states). This is the so called multi-configurations time dependent Hartree-Fock (MCTDHF) approach reviewed in Ref.~\cite{Mey90,Mey09,Hoc11}. 
Such strategy has also been used in Ref.~\cite{Mey90,Man92,Bec00} for bosonic systems  leading to the MCTDH method.
In these cases, the evolution reduces to a set of orthogonal single-particle states whose evolutions are explicitly affected by the evolution of the mixing coefficients $f_\alpha(t)$.
Both MCTDH and MCTDHF equations of motion are greatly simplified due to the fact that the Slater determinants $|\phi_\alpha(t)\rangle$ are orthogonal with each others.

In nuclear physics, a similar multi-configurations approach has, to our knowledge, only been developed in its static version to determine the low energy structure properties of medium nuclei~\cite{Pil08,Rob16,Rob17}.
Meanwhile, many efforts have been spent in considering correlated states built from an ensemble of \textit{non-orthogonal} quasi-particule vacua.
Guided by the specific physical situations (restoration of broken symmetry, large amplitude collective motion, ...), the initial states can generally be labelled by a set of continuous collective coordinates, denoted by $\bm{q}$ such that Eq. (\ref{eq:mix}) usually takes the form:  
\begin{equation}
\label{eq:ansatz0}
 |\psi(t)\rangle = \int_{\bm{q}} f(\bm{q},t) |\phi(\bm{q},t)\rangle d\bm{q}.
\end{equation}
At the initial time, this ansatz encompasses the wide range of situations provided by the static multi-reference framework also referred as Generator Coordinate Method (GCM)~\cite{Rin81}. For instance, in the case of a symmetry breaking/restoration, one would first compute a symmetry breaking mean-field state $|\phi(t_0)\rangle$. From this state, the projection after variation yields a many-body state that can be cast into Eq.~(\ref{eq:ansatz0}) at $t_0$.
In this case, the set of states $|\phi(\bm{q})\rangle$ accounts for the new quasi-particle vacua obtained from the original state $|\phi (t_0) \rangle$ by simple transformations of the symmetry group (e.g. rotation in real and/or gauge space). The mixing function $f(\bm{q},t_0)$ would also be determined up to a non-relevant phase by the value of the quantum number on which we project (e.g. spin, number of particles, ...).
Using such a symmetry breaking/restoration scheme is a key ingredient to describe nuclear correlations with only a few simple many-body states, and is therefore an incentive to develop time-dependent methods based on non-orthogonal quasi-particle vacua.
Applying the same method than for the orthogonal case, it is yet possible to obtain the coupled equation of motion between quasi-particle components forming the many-body vacua and mixing coefficients \cite{Lac19}. 
This results in mean-field like evolutions for the non-orthogonal quasi-particle states that fully couple to each other as well as to the mixing function $f(\bm{q},t)$. The associated numerical effort are anticipated to be significantly increased compared to the standard orthogonal case of Ref.~\cite{Mey09,Hoc11}.

A drastic simplification to this scheme consists in taking a non-variational set of time-independent quasi-particle vacua $|\phi(\bm{q},t)\rangle = |\phi(\bm{q})\rangle$. With such a fixed basis, the variation of the action,  Eq.~(\ref{eq:varia}), is then performed only for the mixing function $f(\bm{q},t)$ and yields the so called Hill-Wheeler equation. This is referred in nuclear physics as the time-dependent GCM approach and is used for instance to understand the fission dynamics~\cite{Ver17}.
The assumption of fixed many-body vacua has clear practical advantages, 
especially for non-orthogonal states, for which a specific treatment of the overlap should be made. 
On the other hand, properly describing the possible exit channels of the dynamics requires a good \textit{a priori} knowledge of the states $|\phi(\bm{q})\rangle$ spanning the corresponding Hilbert subspaces. This pre-selection of states is arbitrary and may directly bias the result. In addition, the basis size itself can become prohibitive when the number of channels increases.

In the present work, we explore an intermediate approximation between the full MC-TDHFB implementation with non-orthogonal states and the extreme assumption
of variation and time independent many-particle vacua (time dependent GCM).
This approach was first proposed in Ref.~\cite{Rei83}.
It starts again from the ansatz given by Eq.~(\ref{eq:ansatz0}) where it is assumed that each of the many-body states $|\phi(\bm{q},t)\rangle$ follows its own TDHF or TDHFB trajectory.
In other words, one first chooses the ensemble of initial quasi-particle vacua $|\phi(\bm{q},t_0)\rangle$. Then the independent TDHFB evolution of each of these states provides the time dependent overcomplete basis $|\phi(\bm{q},t)\rangle$. Finally, the mixing function $f(\bm{q},t)$ is determined from the stationarity of the action Eq.~(\ref{eq:varia}).
This scheme has the great advantage not only to simplify the equations of motion to be solved but also that standard TDHFB solver can be directly used for each individual trajectory.
In addition, we expect that such a time dependent basis will automatically explore the channels beyond the adiabatic ones.
This intermediate implementation of the MC-TDHFB approach is called hereafter \methodi  where ${\rm I}$ stands here for the "Independent" mean-field assumption.

In the present article, we study if this scheme can be used to describe symmetry restoration in time-dependent problems. More specifically, we illustrate it for the case of particle number restoration to describe transfer between two superfluid systems. 

 



\subsection{Equation of Motion of the \methodi}

In this section, we first discuss the general equation of motion to be solved within the \methodi.
In particular, new aspects compared to the case  of time-independent set of vacua are underlined.   
We assume that the initial state is given by Eq. (\ref{eq:ansatz0}) where many-body states have been obtained 
either by symmetry restoration and/or by constrained mean-field techniques. Each of the initial state $|\phi(\bm{q},t_0)\rangle$
is assumed to evolve through its own self-consistent TDHFB equation.  Obviously, in this framework, the proper selection of the initial states will be a determinant ingredient of the accuracy of the approach.  

We apply the variational principle to find the mixing coefficients $f$. A technical but crucial point with the ansatz given by Eq.~(\ref{eq:ansatz0}) is that there is no bijection between $f$ and the many-body state $|\psi(t)\rangle$. Said differently, 
given a set of states $\{ \phi(\bm{q},t) \}$,
there might be several functions $\{ f(\bm q, t) \}$ that lead to the very same state $|\psi(t)\rangle$. This is a well-known 
difficulty linked to the use of non-orthogonal over-complete basis~\cite{Rin81,Rei87}. The problem is usually solved by diagonalizing the overlap kernel $\mathcal{N}$, whose matrix elements are defined as:
\begin{equation}
 \mathcal{N}_{\bm{qq}'}(t) = \langle \phi(\bm{q},t)| \phi(\bm{q}',t)\rangle.  \label{eq:over}
\end{equation}
The situation is slightly more complex when using time-dependent states compared to the case of time-independent states. 
A precise solution to this problem is discussed in appendix \ref{sec:eom}. Ultimately, we have found that the variational principle (\ref{eq:ansatz0}) 
can be safely used by adding the constraint that $f$ belongs to the image of the overlap matrix during the evolution.
With this constraint, we have shown in appendix \ref{sec:eom} that the evolution of the mixing function $f$ can be recast as:
\begin{equation}
\begin{array}{rl}
i\hbar \dot{f} &= \left(\mathcal{N}^{-1} [\mathcal{H} - \mathcal{H}^{MF} ]     
+i\hbar \dot{\mathcal{P}}\right) f,
\end{array}
\label{eq:tdhw_proj}
\end{equation}
where:
\begin{align}
 \mathcal{H}_{\bm{qq}'}(t) &= \langle \phi(\bm{q},t) | \hat{H} | \phi(\bm{q}',t) \rangle, \nonumber \\
 \mathcal{H}^{MF}_{\bm{qq}'}(t) &= \langle \phi(\bm{q},t) | i\hbar \partial_t |\phi(\bm{q}',t) \rangle. \nonumber 
\end{align}
Here, the operator $\mathcal{P}$ corresponds to the projection on the image of $\mathcal{N}$ (in the sense of linear algebra in the space of functions of $\bm{q}$). 
This projector as well as all the kernel operators involved Eq.~(\ref{eq:tdhw_proj}) both depend on time. 
Their explicit  expressions are given in appendix \ref{sec:eom}. 
In addition, this approach contains two new terms compared to the Hill-Wheeler equation usually obtained assuming that quasi-particle vacua do not depend on time.
For time-independent states, only $\mathcal{H}$ enters into the evolution whereas we have here the difference $[\mathcal{H} - \mathcal{H}^{MF}]$ which can be interpreted as 
the residual coupling between states that is not already accounted for in the mean-field evolution of the states.
This is a nice aspect of using the variational principle (\ref{eq:varia}) that automatically avoids double-counting in the evolution 
between the mixing coefficients and the waves evolutions. Said differently, the coefficients evolution are automatically adjusted not to 
account for the information already contained in the state evolution. Another difference is the presence of the $\dot{\mathcal{P}}$ term.
It is a direct consequence of imposing that $f$ stays in the image of $\mathcal{N}$ at all time.  
This contribution is directly linked to the overlap matrix evolution and disappears for time-independent states. 

Starting from an initial mixed state, one could numerically integrate Eq.~(\ref{eq:tdhw_proj}) to obtain the dynamics of the system. However, it 
has many practical interest to transform it into another one for the so--called collective wave-function that we note here $g(\bm{q},t)$~\cite{Rin81,Rei87} and that is defined as:
\begin{equation}
 g = \mathcal{N}^{1/2}f.
\end{equation}
Following a standard procedure, we also transform any observable kernel $\mathcal{O}$ to its collective operator $\mathcal{O}^c$:
\begin{align}
 \mathcal{O}^c &= \mathcal{N}^{-1/2} \mathcal{O} \mathcal{N}^{-1/2}.
\end{align}
We recall that the benefit of such representation is to have an isomorphic mapping of the observables which brings out handy properties like for instance:
\begin{align}
 \langle \psi(t) |\hat{O} | \psi(t) \rangle &= \langle g | \mathcal{O}^c | g\rangle, \nonumber \\
 ([\hat{A},\hat{B}])^c &= [\mathcal{A}^c, \mathcal{B}^c] , ~~~
 (\hat{O}^\dagger)^c = (\mathcal{O}^{c})^\dagger.  \nonumber
\end{align}
The equation of evolution for the collective wave-function directly derives from Eq.~(\ref{eq:tdhw_proj}) and becomes:
\begin{equation}
\begin{array}{rl}
 i\hbar \dot{g} &= \left(\mathcal{H}^c -\mathcal{H}^{MF,c} + i\hbar \dot{\mathcal{N}}^{1/2}\mathcal{N}^{-1/2} \right)g.
\end{array}
\label{eq:tdhw_proj_g}
\end{equation}
This is the equation that we use in practice in the following applications.  


To summarize, the essence of the \methodi approach is to replace 
the problem of many TDHFB coupled equations by two independent problems. The first one is to propagate 
in time a set of independent mean-field trajectories. The second one is to propagate in time the collective wave function according to Eq.~(\ref{eq:tdhw_proj_g}). This  requires the determination of a few collective kernels at any time of the evolution. 
In practice, compared to the case of time-independent state, this approach trades off a reduction of the number of basis states at the price of an increased complexity due to the time-evolution of the overlap kernels. Some first tests of this approach were performed in Ref.~\cite{Rei83} based on additional approximations for the estimation of the kernels involved in Eq.~(\ref{eq:tdhw_proj_g}) (e.g. Gaussian overlap approximation).
In what follows, we investigate the full fledged approach to describe the particle transfer during a 
short contact between two small superfluid systems.

\section{Description of the transfer of particles between two superfluid in a model case}
\label{sec:application}

To assess the performance of the \methodi approach, we now consider two small superfluid systems that interacts with each 
other over a short time interval and then re-separate.
We use the same model as in our previous work~\cite{Reg18} where the two
systems labeled by $A$ and $B$ respectively are represented by $\Omega_A$ (resp. $\Omega_B$) single-particle energy levels. Each level is doubly degenerated and may contain zero or one pair of fermions (see also \cite{Die70,Die71,Har71}).
The total Hamiltonian is a sum of two pairing Hamiltonian~\cite{Bri05} acting on $A$ and $B$ respectively, 
plus a coupling term that enables the exchange of pairs. 
The complete Hamiltonian can be written as: 
\begin{eqnarray}
H  =  \sum_{k\in A\cup B } \varepsilon_k (a^\dagger_k a_k + a^\dagger_{\bar k} a_{\bar k}) 
- \sum_{ k,l \in A\cup B}  g_{kl} a^\dagger_{ k} a^\dagger_{\bar k} a_{\bar l} a_l,
\label{eq:hab_bis}
\end{eqnarray} 
with the shorthand notation for the pairing matrix
\begin{align}
 g_{kl} = \left\{
 \begin{array}{ll}
 g & \text{if }kl \in A \text{ or } kl\in B, \nonumber \\
 v(t) & \text{if }k\in A, l \in B \text{ or } k\in B, l \in A.
 \end{array}
 \right.
\end{align}
In Eq. (\ref{eq:hab_bis}), we use a generic notation $\{ a^\dagger_k, a_k\}$ for creation/annihilation of single-particle states belonging either to  the 
sub-system $A$ or $B$. 
The time dependency of the coupling term $v(t)$ simulates a short contact between the two superfluids and we adopted a simple Gaussian shape.

\begin{equation}
 v(t) = v_0 \, \text{exp}(-t^2/\tau_c^2).
\end{equation}
We take here a characteristic time $\tau_c=0.28 \hbar/g$ and will look at the evolution in the time range $[-1.2, +1.2] \hbar/g$. The first part of dynamics up to $t~\simeq -0.6$ basically corresponds to the evolution of the two uncoupled systems. Then an interaction of varying strength takes place up to $t=~0.6 \hbar/g$. Finally we have again a phase of uncoupled evolution. 
In the case of nuclear physics this situation typically mimics the collision between two open-shell nuclei below the Coulomb barrier.

One interest of the present
model is that it can be solved exactly giving a stringent test of possible approximate treatment of the evolution. This model was recently 
used for instance to illustrate the validity of standard TDHFB dynamics and the dependency of particle transfer with the initial relative 
gauge-angle in Refs. \cite{Sca17,Sca17b,Reg18}. In these studies, 
the possibility to go beyond the mean-field picture was explored. 
Both approaches have however pointed out some difficulties in reproducing 
the exact evolution. In \cite{Sca17,Sca17b}, a method was proposed to perform the mixing of several TDHFB trajectories. However, it was shown 
that the approximation made on the relative phase between trajectories strongly influences the predictive power. For this reason, a semi-classical phase-space approximation was studied in ref. \cite{Reg18} where a statistical mixing of TDHFB trajectories was assumed. 
The conclusion was that a good reproduction can only be obtained by 
an artificial rescaling of the coupling interaction together with a careful analysis of the probabilistic interpretation of the result.    
Anticipating the discussion below we show that the MC-TDHFB$_{\textrm{I}} $ method does not face this difficulty.

We concentrate here on the collision between two superfluid systems. In the illustration presented below, we will always be in situations where 
the pairing strength $g$ appearing is sufficiently high to insure that the solution of the HFB 
equation for each subsystem ($A$ and $B$) 
gives quasi-particle vacuum breaking the particle number symmetry. At the HFB level, the ground state is determined up to a phase in the 
pairing field and the TDHFB equations are solved for the composite system [Eqs. (13) or [(14)] in Ref. \cite{Reg18}]. 

At the MC-TDHFB$_{\textrm{I}} $ level, a set of independent TDHFB 
evolutions will be performed and their quantum mixing will be obtained by solving Eq. (\ref{eq:tdhw_proj_g}). In practice, we need 
first to specify the initial conditions (choice of states, number of states and initial mixing).   
 
\subsection{Initial condition for the MC-TDHFB$_{\textrm{I}} $ method}
\label{sec:initial}

In the HFB theory, the many-body state is composed initially by two quasi-particle vacua describing the systems $A$ and $B$. 
This state corresponds to a mixture of many-body states with different particle numbers. 
As a consequence, in the calculated observables, there is an inherent contribution from 
particles numbers that differ from the one under interest. A natural way to remove this spurious contribution is to project the wave function 
onto the desired particle number \cite{Rin81}. In the physical situation we are interested in, initial subsystems have definite
particle numbers equal to $N^0_A$ and $N^0_B$ respectively.        

In this application, we first determine the HFB ground state $|\varphi_A(t_0)\rangle$ (resp. $|\varphi_B(t_0)\rangle$) of the uncoupled system $A$ (resp. $B$)
\begin{equation}
 |\varphi_{A/B}(t_0) \rangle = \prod_{i\in {A/B}} (u^*_i + v^*_i a^\dagger_{i}a^\dagger_{\bar{i}})|0_{A/B}\rangle,
\end{equation}
where the $u_i,v_i$ are the quasiparticles components for each level $i$ of the subsystem. $|0_A\rangle$ and $|0_B\rangle$ denotes here the particle vacuum 
of sub-systems $A$ and $B$ respectively.  
Then, we build the antisymmetric HFB product state
\begin{eqnarray}
|\phi(t_0)\rangle= 
\prod_{i\in A\cup B}(u^*_i + v^*_i a^\dagger_i a^\dagger_{\bar{i}})
|0_A\rangle \otimes |0_B\rangle
\label{eq:compound}
\end{eqnarray}
for the compound system. 
This state is the one that is usually used as initial condition to treat nuclear collisions within TDHFB~\cite{Has16}. 

A state with proper particle number is obtained by projecting the symmetry-breaking state (\ref{eq:compound}) both on the total number of particle $N^0 = N^0_A + N^0_B$ and on the desired particle number of one of the sub-system
($A$ in the following).    
This procedure can be written in a compact form as:
\begin{eqnarray}
|\psi(t_0)\rangle  = {C} \hat{P}(N^0) \hat{P}_{A}(N^0_A) |\phi(t_0)\rangle.  \label{eq:pp}
\end{eqnarray}     
Here $C$ is a constant that insures the normalization of the state. Here, $\hat{P}(N^0)$  and $\hat{P}_{A}(N^0_A)$ denote the projectors on the total number of particles $N^0$
and on the number of particles $N^0_A$ in $A$ respectively. Explicit forms of the projectors in terms of integrals on gauge angle can be found in Ref.  \cite{Rin81}. 
Note that Eq. (\ref{eq:pp}) is strictly equivalent to project  each subsystem separately on their respective particle number. 
However, as we will see, the use of the projector on total particle number has some practical advantage. It will indeed enable us later on to leverage the conservation of the total number of particles to reduce the numerical complexity of the problem.
Writing down the projectors explicitly as a mixture of rotations in the gauge space, we obtain for the initial state:
\begin{equation}
\label{eq:istate}
 |\psi(t_0)\rangle = \int_{\theta=0}^\pi \int_{\theta_A=0}^{\pi} f(\bm{\theta},t_0) |\phi( \bm{\theta},t_0)\rangle d\bm{\theta},
\end{equation}
with
\begin{align}
|\phi(\bm{\theta},t_0)\rangle = |\phi(\theta,\theta_A,t_0)\rangle 
&= e^{i\theta(\hat{N}-N^0)} e^{i\theta_A(\hat{N}_A-N^0_A)} |\phi(t_0)\rangle, \nonumber \\
f(\bm{\theta},t_0) &= \frac{1}{\pi^2 \sqrt{P(N^0,N_ A^0)}}.
\end{align}
Here the operators $\hat{N},\hat{N}_A$ are associated to the total number of particles and the number of particles in $A$ respectively, whereas $P(N^0,N_A^0)$ is the probability to measure $N^0$ total particles and $N^0_A$ particles in $A$ from the HFB state $|\phi(t_0)\rangle$. With these conventions, the function $f$ is a constant independent on $(\theta, \theta_A)$, that can be determined by the normalization condition on the initial state. In practice, we used the Fomenko scheme~\cite{Fom70} to discretize the projectors which implies that the angles in gauge space take the values $l\pi/L$ for the integer $l\in[1,L]$.

\subsection{Solving the dynamics at the MC-TDHFB$_{\textrm{I}} $ level}

We give here some specific aspects related to the time-dependent configuration mixing method applied to particle number 
symmetry restoration.    
The first ingredient to determine the MC-TDHFB$_{\textrm{I}} $ dynamics is the computation of the independent TDHFB evolutions $|\phi(\bm{\theta},t)\rangle$ starting from the rotated states in gauge space. For more details on this standard calculation, we refer the reader to our previous work~\cite{Reg18}.
On top of this, we numerically integrate the collective dynamics given by Eq. (\ref{eq:tdhw_proj}).
At each time step, this implies the calculation of different kernel operators.
The overlap matrix between two states with different gauge angles reads
\begin{align}
\label{eq:overlap}
 \mathcal{N}_{\bm{\theta}\bm{\theta}'}(t) 
 &= \prod_{i} \left[u_i^{\bm{\theta}} u_i^{\bm{\theta}'*} + v_i^{\bm{\theta}} v_i^{\bm{\theta}'*} \right], 
\end{align}
where the $u^{\bm{\theta}},v^{\bm{\theta}}$ are the quasi-particles components of the state $|\phi(\bm{\theta},t)\rangle$.
Similarly, the Hamiltonian kernel is the sum of a 1-body and 2-body part:
\begin{align}
\label{eq:hkernel}
%
\mathcal{H}^{(1)}_{\bm{\theta}\bm{\theta}'} &=  
\sum_{k} 
(2\varepsilon_k - g_{kk}) v_k^{\bm{\theta}} v_k^{\bm{\theta}'*} \prod_{i\neq k}(u_i^{\bm{\theta}} u_i^{\bm{\theta}'*} + v_i^{\bm{\theta}} v_i^{\bm{\theta}'*}) , \nonumber \\
\mathcal{H}^{(2)}_{\bm{\theta}\bm{\theta}'} &=  
  - \sum_{\substack{k,l \\ k\neq l}} g_{kl}(u_l^{\bm{\theta}} v_l^{\bm{\theta}'*})( u_k^{\bm{\theta}'} v_k^{\bm{\theta}*})^*
 \prod_{i\neq kl} (u_i^{\bm{\theta}} u_i^{\bm{\theta}'*} + v_i^{\bm{\theta}} v_i^{\bm{\theta}'*}).
\end{align}
Finally we need the time derivative kernel (or the mean-field Hamiltonian kernel).
\begin{align}
\label{eq:mfkernel}
  \mathcal{H}^{MF}_{\bm{\theta}\bm{\theta}'} 
 &=  i\hbar\sum_{k}  
 (u_k^{\bm{\theta}} \dot{u}^{\bm{\theta}'*}_k + v_k^{\bm{\theta}}  \dot{v}^{\bm{\theta}'*}_k )
  \prod_{i\neq k} (u_i^{\bm{\theta}} u_i^{\bm{\theta}'*} + v_i^{\bm{\theta}} v_i^{\bm{\theta}'*}).
\end{align}
This effective hamiltonian involves time derivatives of the $u^{\bm{\theta'}},v^{\bm{\theta}'}$ components that we estimate in practice by using the TDHFB evolution equation 
associated to each trajectories.
Once these kernels are known, the overlap kernel is explicitly diagonalized so to determine $\mathcal{N}^{-1/2}$ as well as the collective operators.
Finally, the time derivative $\dot{\mathcal{N}}^{1/2}$ is estimated numerically using a finite difference scheme.
With this, we integrate Eq.~(\ref{eq:tdhw_proj_g}) for a short time step using a fourth order Runge-Kutta scheme.
Repeating this procedure yields the complete evolution of the collective wave function $g(\bm{\theta},t)$.

\subsection{Leveraging global symmetries in the \methodi dynamics}

In the present work, we apply the \methodi  to the problem of restoration of broken symmetries in dynamical problem. In general, 
the projection of the complete system are made by considering symmetry-broken states that are obtained one from another by 
specific transformations (rotation in 3D space for the rotational invariance, rotation in the $U(1)$ space for the particle number). 

\begin{figure}[ht!]
 \includegraphics[width=0.8\linewidth]{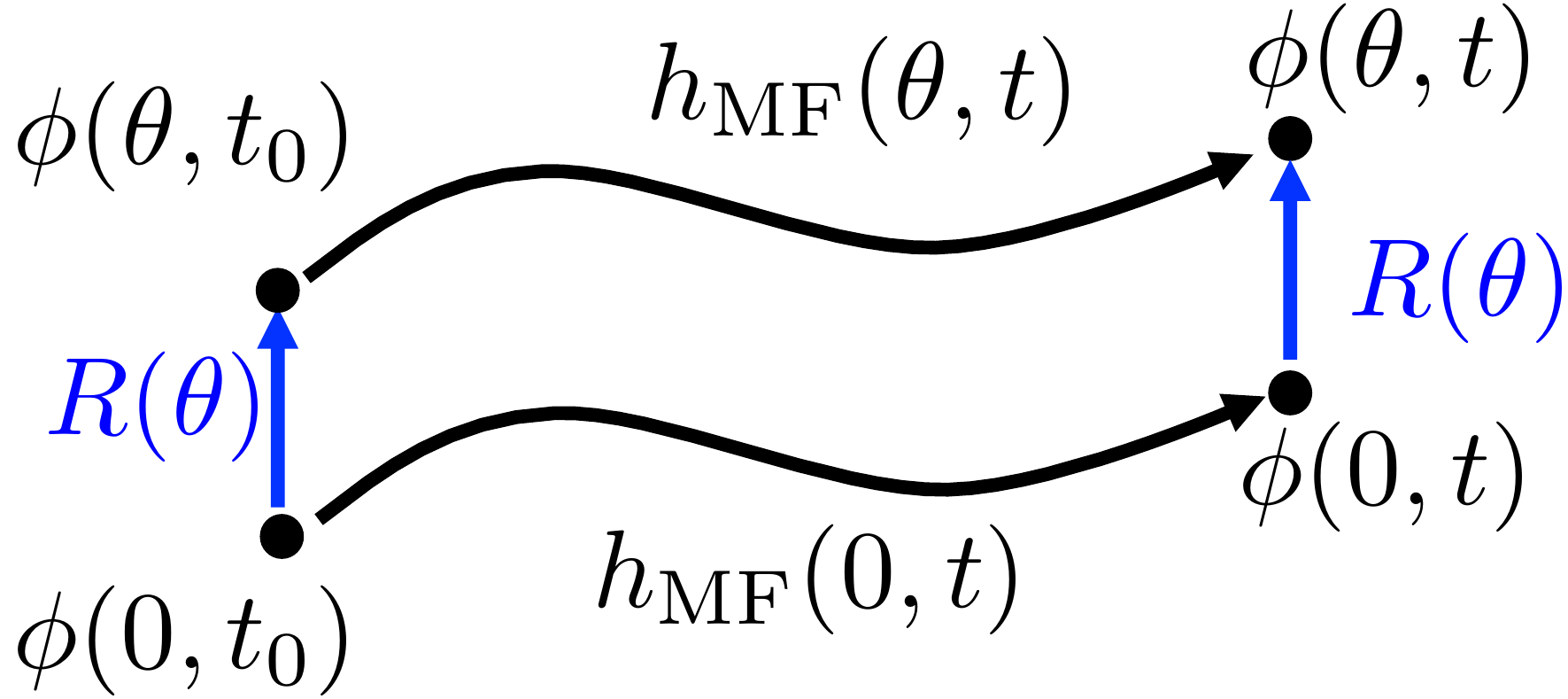}
 \caption{Illustration of the fact that it is equivalent to first rotate (through a rotation operator $R(\theta)$) a mean field state at $t_0$ and then evolve it with its mean-field propagator, or perform its mean-field evolution first and then rotate the state at time $t$.}
 \label{fig:rotglobal}
\end{figure}

Let us first consider the case of rotation in real space that is easier to visualize. Let us assume a system that is deformed initially at the mean-field level and is associated to an initial 
vaccum (called here original state). We then perform its evolution. The restoration of the total angular momentum in the \methodi is equivalent to mix up different initial    
conditions obtained from the original state through a rotation. Since the rotation is a global rotation of the whole system, the evolution of the rotated state turns out to be the same 
as the original evolution in a different frame (this frame being fixed in time and corresponds to the frame obtained by rotation of the original frame). 
A corollary to this simple argument is that the state obtained by evolving the initially rotated state with its own mean-field at a given time $t$ also corresponds to  
the state obtained by considering the original state evolved up to time $t$ also with its own mean-field followed by a rotation. Similar conclusion can be drawn here for the rotation in gauge-space. This specific case is illustrated in Fig. \ref{fig:rotglobal}.
Note however that 
this only applies if a transformation of the entire system is made, i.e. a global symmetry is restored. This is not the case, if one rotates   
only part of the system for instance to make the projection on $N^0_A$.

To illustrate how this can simplify the application of \methodi, let us assume first that we make a single projection on the total particle 
number $N^0$.  Then, the initial state writes:
\begin{equation}
 |\psi(t_0)\rangle = \int_{\theta=0}^\pi 
 f(\theta,t_0) |\phi( \theta,t_0)\rangle d{\theta},
\end{equation}
where $|\phi(\theta,t_0)\rangle \equiv  e^{-i\theta N^0} \hat R(\theta) |\phi(\theta=0,t_0)\rangle$. $\hat R(\theta) = e^{i\hat{N}\theta}$ denotes the rotation operator in gauge space. 
 At a given time $t$, we have:
\begin{equation}
 |\psi(t)\rangle = \int_{\theta=0}^\pi  
 f(\theta,t) |\phi( \theta,t)\rangle d{\theta},
\end{equation}
where $|\phi( \theta,t)\rangle$ are the initially rotated states evolved with their own mean-field while $ f(\theta,t)$ are obtained through the coupled equations (\ref{eq:tdhw_proj}). 
According to the discussion made above, there is no need to perform the mean-field evolution of each initially rotated state and we simply can use:
\begin{align}
\label{eq:rot_on_mf}
|\phi(\theta,t)\rangle &=  e^{-i\theta N^0} \hat{R}(\theta)|\phi(\theta=0,t)\rangle.
\end{align}
This means in practice that, in the case of a single projection on the total particle number, we only need to follow one state (the non-rotated one)
and not a full set of $\theta$ dependent mean-field. This already is a great simplification. 

We now consider the case where we both project initially on the total particle number $N^0$ and on the number of particle $N^0_A$  in system $A$. 
We assume that the two projections are associated to the gauge-angle $(\theta, \theta_A)$. We can now make use of the fact:
\begin{align}
|\phi(\theta,\theta_A,t)\rangle = e^{-iN^0\theta}\hat{R}(\theta)|\phi(0,\theta_A,t)\rangle.
\end{align}
This relation enables in practice to perform only the TDHFB evolutions associated with $\theta=0$, reducing by a factor $L$ (the number of discrete $\theta$ angles) the cost associated to these independent propagations.

\begin{figure}[ht!]
 \includegraphics[width=0.95\linewidth]{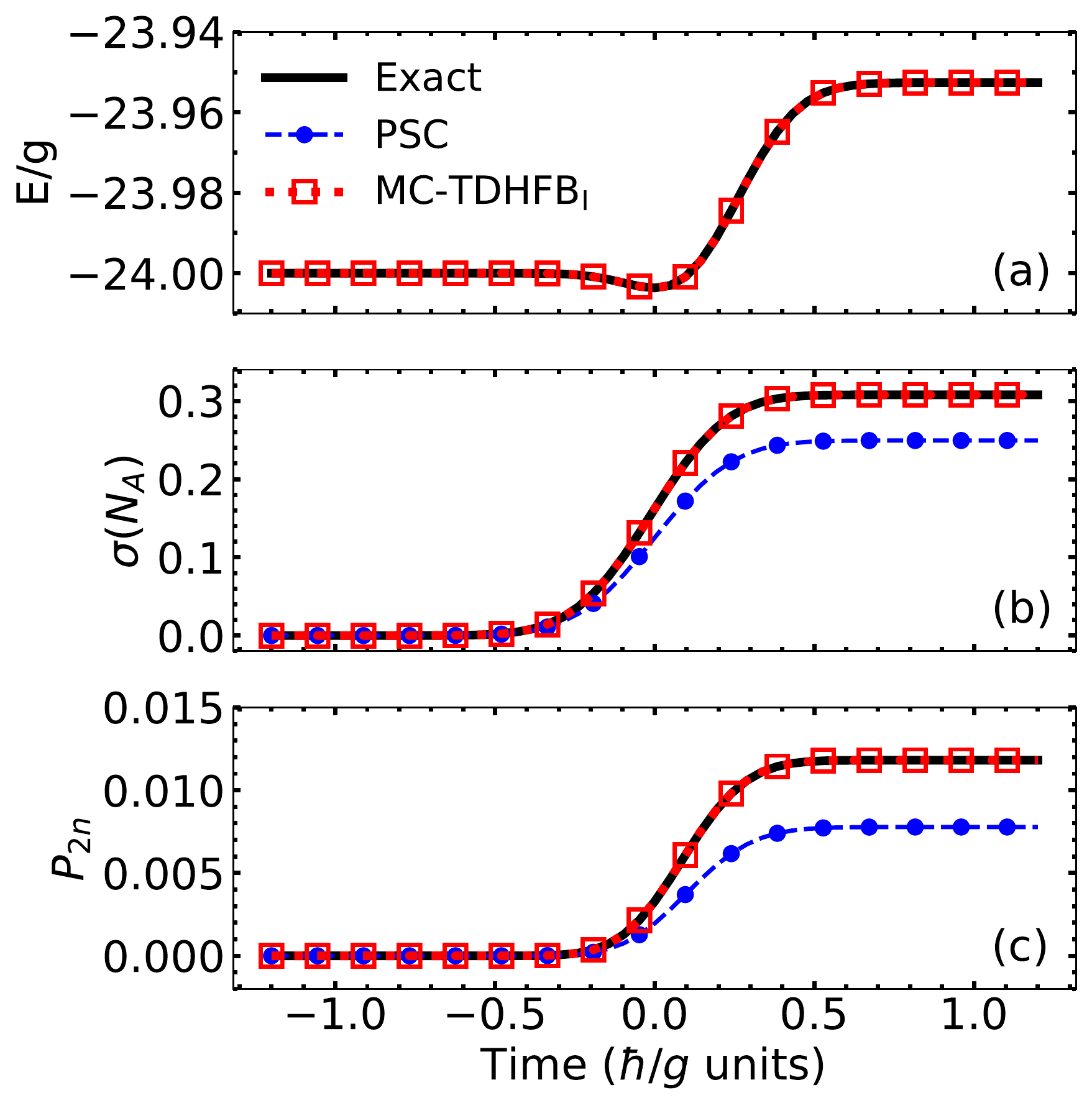}
 \caption{Evolution of the total energy (a), fluctuation in particle number $A$ (b) and the probability of one pair transfer $P_{2n}$ (c) as a function of time. Calculations were performed with a coupling strength $v_0=0.02g$ (perturbative regime) for the symmetric case  $\Omega_A=\Omega_B=6$ and $N_A^0=N_B^0=6$.  
 The \methodi approach (red squares) is compared with the exact solution (black line) and the the phase-space approach of Ref.~\cite{Reg18} without renormalization of the interaction (blue filled circles).}
 \label{fig:allobs}
\end{figure}

\subsection{Collision between two identical and degenerated systems}

To illustrate the MC method, we consider first the collision between two identical systems with $\Omega_A=\Omega_B=6$ and $N_A^0=N_B^0=6$. We assume 
in addition that each system is fully degenerated, i.e. $\varepsilon_k = 0$ for all the levels $k$. 
For this system, we performed a series of MC-TDHFB$_{\textrm{I}} $ evolutions and considered different coupling strengths between the two systems 
from the very perturbative regime to highly non perturbative cases. Each of these calculations was repeated with different time steps and dimension $L$ for the discretization of the projectors. We found that the results properly converge for $dt = 10^{-3}$ $\hbar/g$ and 7 discrete angles.

Fig.~\ref{fig:allobs} shows the total energy as a function of time as well as the probability $P_{2n}$ for the transfer of one pair of particle from $A$ to $B$ for a coupling strength $v_0= 2\times10^{-2}\ g$. Such a probability is obtained by projecting the finale state on the good particle number in the subsystem $A$. This $v_0$ value corresponds to a perturbative regime where the final probability for the transfer of one pair during the process is $P_{2n}\simeq1.2\times10^{-2}$.
We see that MC-TDHFB$_{\textrm{I}} $ reproduces exactly both the fluctuation $\sigma(N_A)$ and the total energy at any time. For comparison, we computed the same quantities with the phase-space combinatorial (PSC) approach developed in Ref.~\cite{Reg18} but without any renormalization of the coupling $v_0$. 
We recall that this method is based on the same TDHFB trajectories but the observables are computed as classical average over the mean-field expectation values in the PSC case. 
This semi-classical approach reproduces pretty well the exact fluctuations with still a 20\% underestimation at the final time. This discrepancy translates the effect of the quantum interferences between the TDHFB trajectories that is well accounted for in MC-TDHFB$_{\textrm{I}} $ but is not in the PSC.
%

The fact that the MC-TDHFB$_{\textrm{I}} $ reproduces so well the exact solution holds for any coupling constant $v_0$ and may appear disturbing at first glance. It is actually explained by the fact that the relevant Hilbert space for fully degenerated subsystems is very small.
Due to the degeneracy of the levels in $A$ and $B$, there is only one possible state $|N_A,N_B\rangle$ for each couple of values 
($N_A$, $N_B$) that plays a role in the transfer process. 
In addition, we know that the total number of particles in the system is conserved by the Hamiltonian. This implies that the exact solution belongs to the subspace $E(N=6)$ of dimension 7 spanned by the states $|N_A=n, N_B= 6 - n\rangle$ with $n\in [0,6]$.
It is therefore sufficient that the TDHFB trajectories $|\phi(\bm{\theta},t)\rangle$ projected on the good number of particles span a space of dimension $7$ at any time to ensure that the MC-TDHFB$_{\textrm{I}} $ solution is the exact one. 
We performed calculations for the coupling constants $v_0= 2\times10^{-3}g,\ 2\times10^{-2}g,\ 2\times10^{-1}g$ and $2g$ and found that this criteria  is always met at any time of the evolutions.
In this simple fully-degenerated situation, the MC-TDHFB$_{\textrm{I}} $ approach therefore provides the exact solution.
\begin{figure}[t!]
 \includegraphics[width=0.95\linewidth]{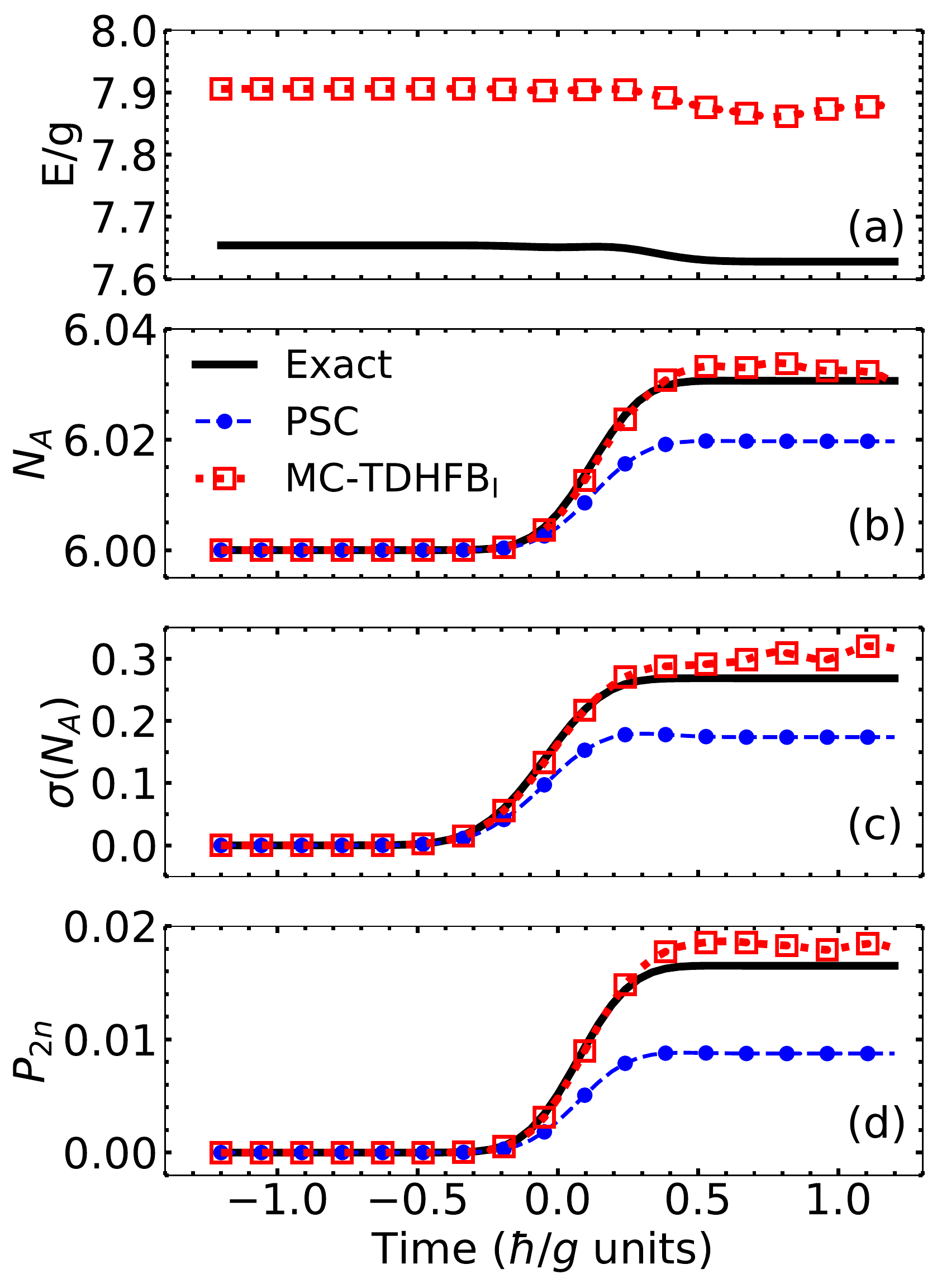}
 \caption{Evolution of the energy (a), the number of particles in $A$ (b), its fluctuation (c), and the probability of one pair transfer from $B$ to $A$ (d) as a function of time for asymmetric collisions. Calculations are performed in a perturbative regime for the coupling strength $v_0=2\times 10^{-2} g$.  The 
 MC-TDHFB$_{\textrm{I}} $ approach (red squares) is compared with the exact solution (black line) and the phase-space approach of Ref.~\cite{Reg18} without renormalization of the interaction (blue filled circles). The energy predicted with PSC (not shown) evolves in the range [11.44,11.54] $g$.}
 \label{fig:allobs_asym}
\end{figure}
%
\begin{figure}[t!]
\includegraphics[width=0.95\linewidth]{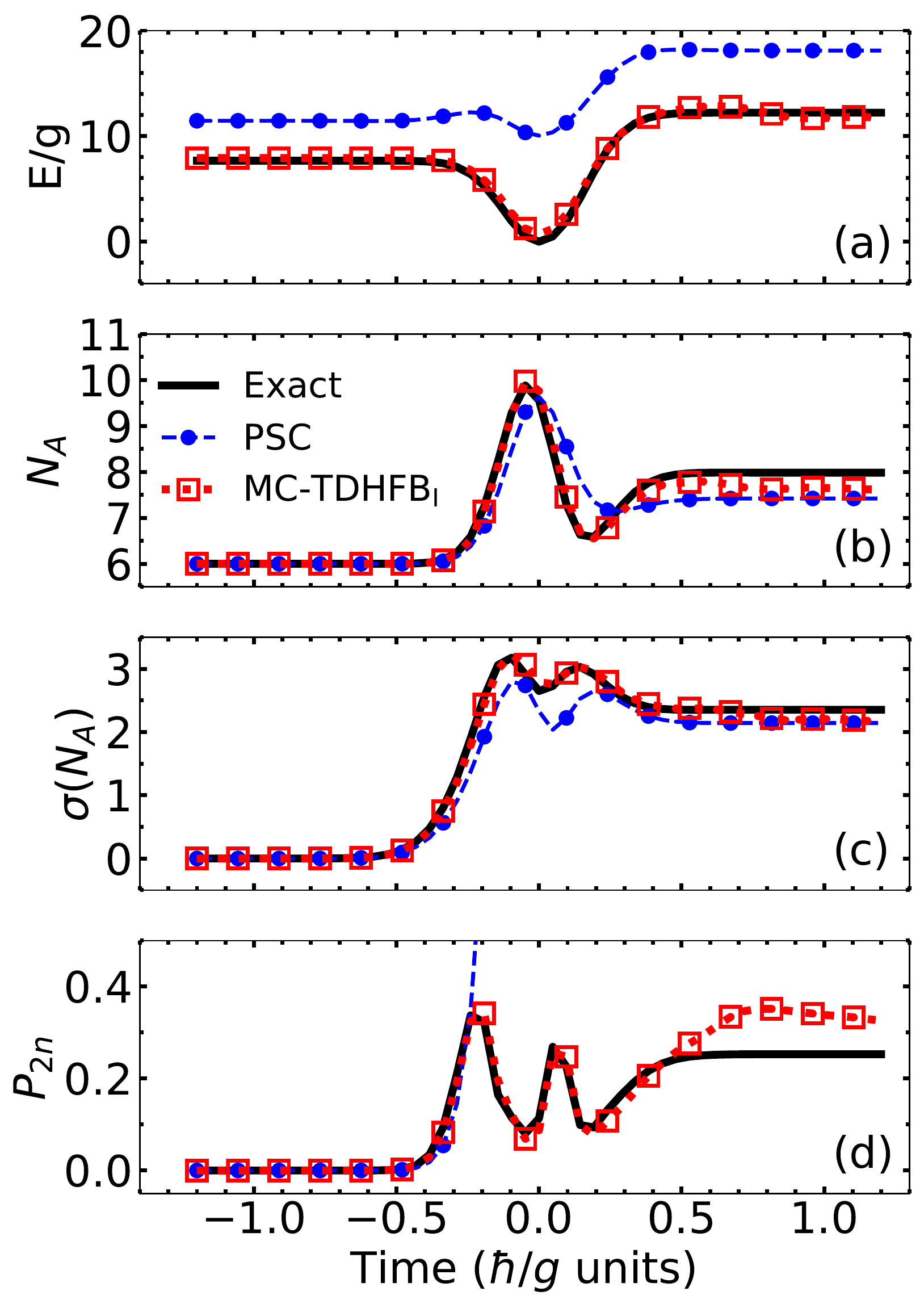}
 \caption{Same as Fig.~\ref{fig:allobs_asym} for the non perturbative regime with $v_0=  g$.}
 \label{fig:allobs_asymnp}
\end{figure}
\subsection{Collision between different non-degenerated system}

Let us now consider a more complex situation where the two systems are characterized by $\Omega_A=\Omega_B=8$ and are filled with the different particle numbers $N_A^0=6$ and $N_B^0=10$. In addition we assume non-degenerated single-particles levels in both subsystems by choosing $\varepsilon_k = kg$ with $k\in[0,7]$. In this configuration, the number of states with the good total number of particles is 12870.
The energy splitting between levels yields a superfluid phase for the HFB ground states of the two subsystems. Their associated occupation numbers are in the range $n_k\in [0.07, 0.90]$ which corresponds typically to the situation of active nucleons around the Fermi surface in mid-shell nuclei.
To converge the solution, we took a smaller time step than for the previous case, $dt=3\times10^{-4} \hbar/g$. Finally, we present the results obtained with $L=25$ angles to discretize the $\theta$ range except if we explicitly specified otherwise.

When the two systems enter into contact, they first tend to equilibrate their number of particles. The  rapidity of this equilibration and the possible presence of oscillations around the half-filling configuration ($N_A=N_B=8=\Omega_{A/B}/2$) depend on the interaction strength $v_0$.
%

We show in Fig.~\ref{fig:allobs_asym} the results obtained in a perturbative regime with $v_0=0.02 g$.
We see in panel (a) that the energy follows the same trend as the exact solution with a small constant shift in energy between the exact energy and the \methodi energy. This shift, already present at initial time, simply illustrates the fact that the initial state used in the configuration-mixing approach (see section \ref{sec:initial}) does not grasp all the correlations present in the exact ground state of the separated systems.
In this regime, we also see in panel (b) that the average number of particles in $A$ increases monotonically by up to 0.7\% of its initial value. 
Due to the weak coupling, this increase is rather small compared to the full chemical equilibration between the two sub-systems.
The \methodi reproduces the exact value for the average drift, the fluctuation of $N_A$ and the probability $P_{2n}$ to transfer one pair from $B$ to $A$ within 10\%.
The phase-space combinatorial approach is also given for the sake of comparison and we see similar behavior as was already mentioned in the symmetric case. Once again, the deviation between PSC and \methodi can directly be attributed to the quantum interferences between mean-field trajectories. We see that this interference effect is responsible for basically half of the probability to transfer one pair of particle from $B$ to $A$ in the perturbative regime. The quality of the \methodi prediction for $P_{2n}$ as well as the contribution of the interferences remains the same in the whole perturbative regime.
\begin{figure}[ht!]
 \includegraphics[width=0.95\linewidth]{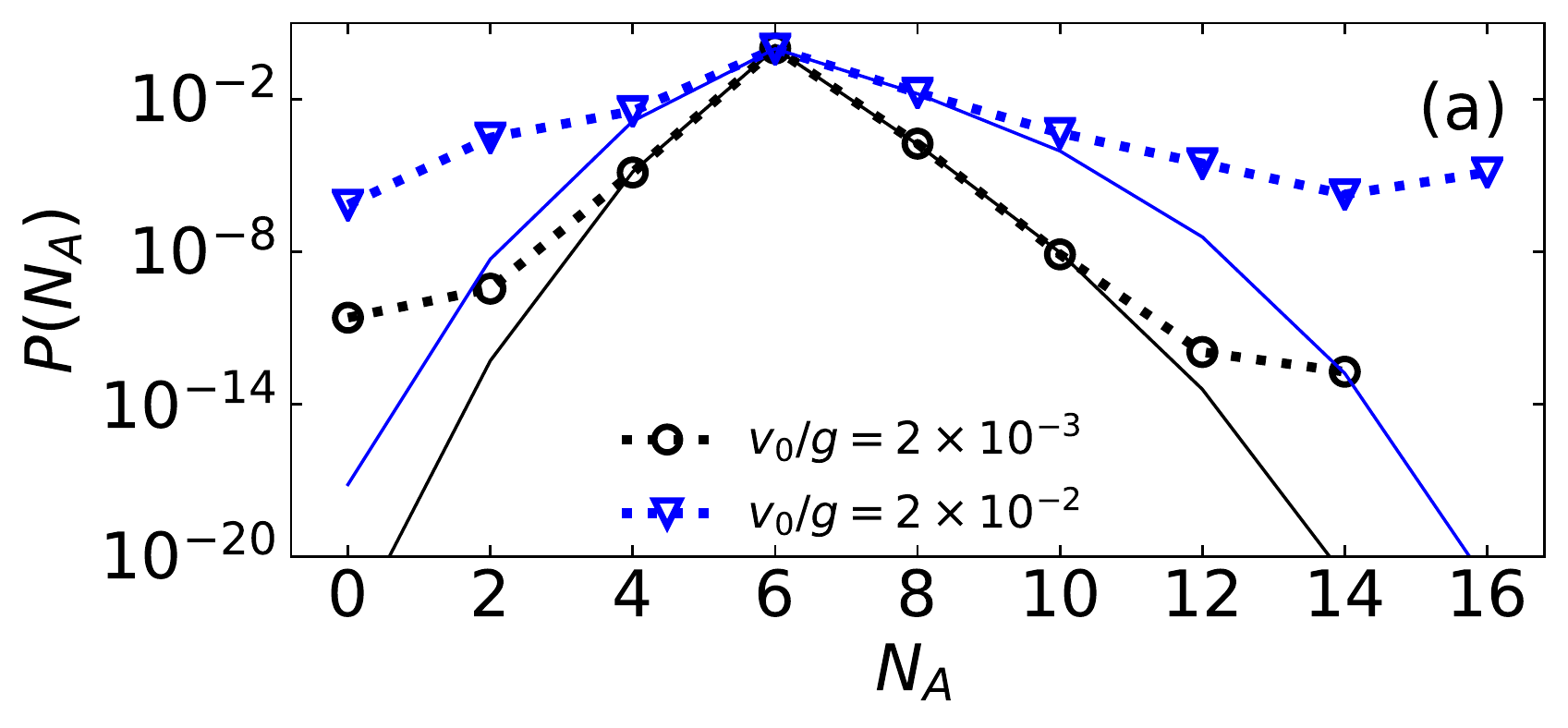}
 \includegraphics[width=0.95\linewidth]{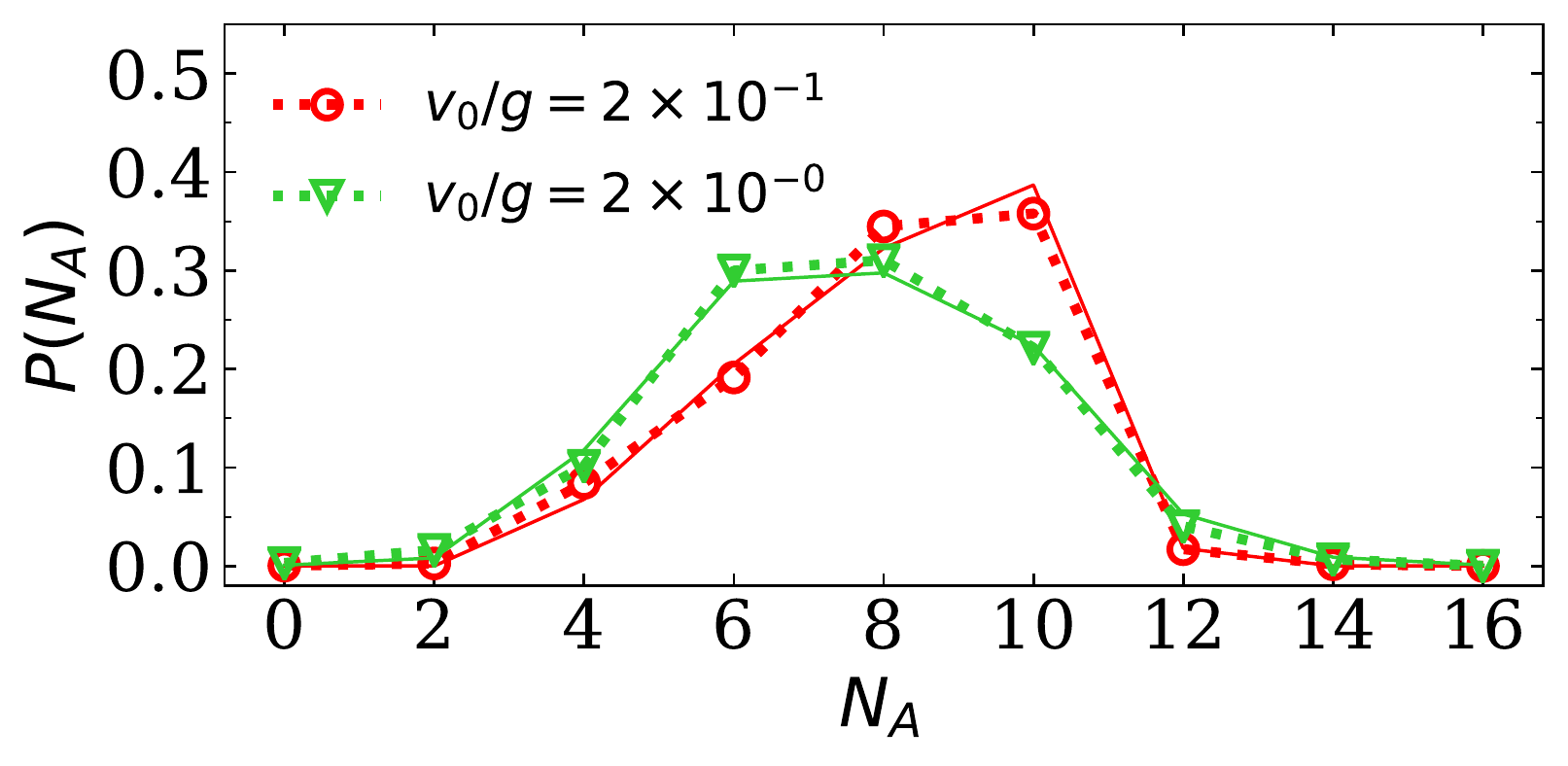}
 \caption{Distribution of probabilities of the final number of particles in the system $A$. The probabilities are calculated after the collision between two non-degenerate systems with $\Omega_A=\Omega_B=8$ and $N_A^0=6$, $N_B^0=10$. Calculations are repeated for several coupling strength $v_0$. (a) Distribution of probability for the low coupling strengths represented in a logarithmic scale. (b) Same probability distribution for  stronger coupling strengths. Here a linear scale is used for the $y$-axis.
 }
 \label{fig:n1distrib_asym}
\end{figure}
%


We performed similar calculations in the non perturbative regime and show the results for $v_0=g$ in Fig.~\ref{fig:allobs_asymnp}. 
As discussed in details in our previous work~\cite{Reg18}, the PSC approach fails to predict the one pair transfer probability in this regime where the multi-pair transfer channels play a non negligible role.
In comparison, the \methodi approach reproduces the energy, average drift and fluctuation of $N_A$ and the probability of one pair transfer with similar quality as in the perturbative regime. Correctly taking into account the interferences in the collective dynamics, we gained from PSC the ability to predict the one pair transfer in the complete range of coupling strength.

\begin{figure}[ht!]
 \includegraphics[width=0.95\linewidth]{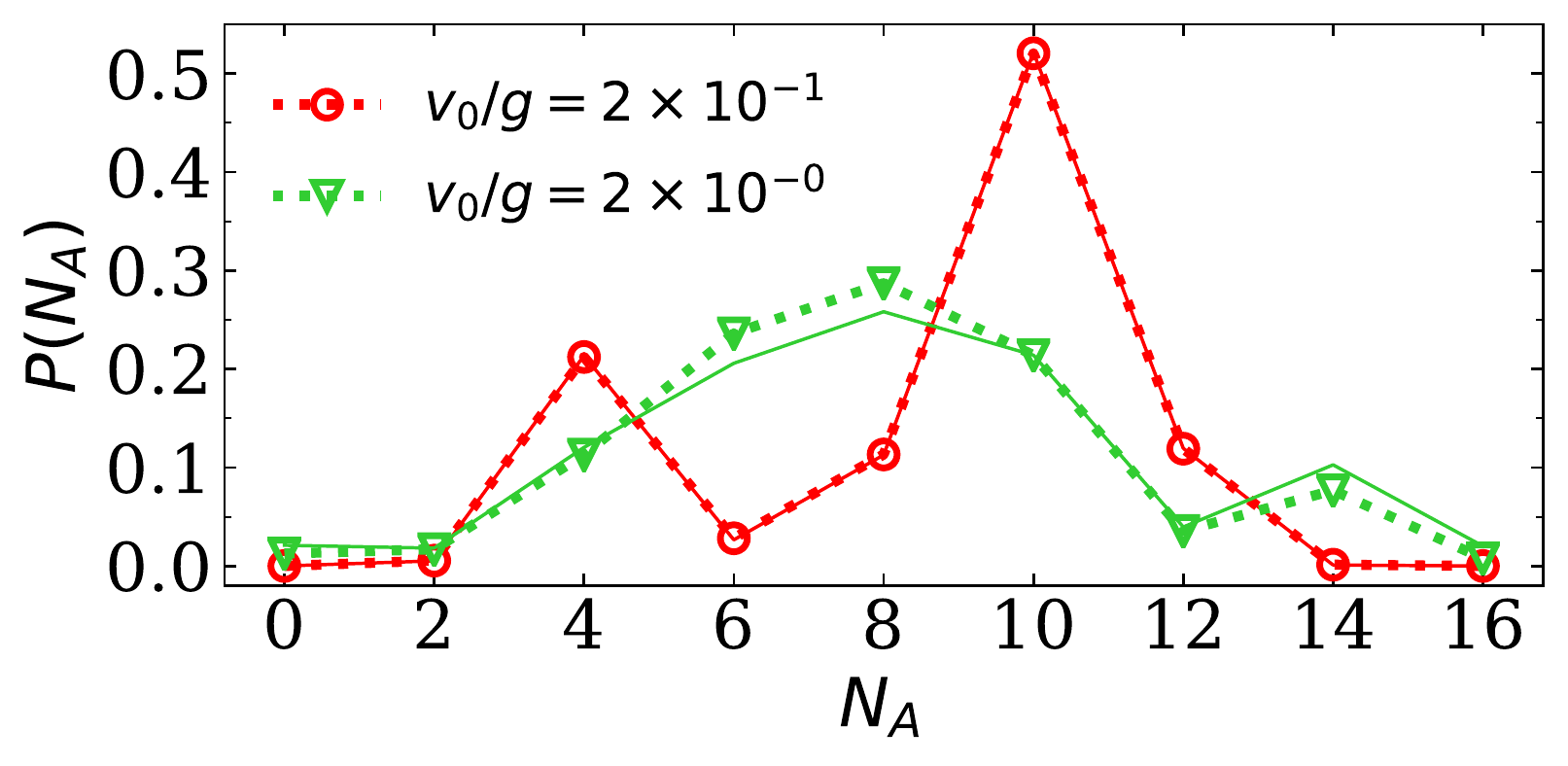}
 \caption{Same as Fig.~\ref{fig:n1distrib_asym} pannel (b) when the energy of the single particle levels is set to $\epsilon_k=kg/4$.}
 \label{fig:n1distrib_dE0p25}
\end{figure}

Another benefit of the \methodi approach  is the possibility to determine the multi-pair transfer probabilities directly from the final wave-function. In practice, this 
is achieved by projecting out the strongly entangled state given by Eq. (\ref{eq:ansatz0}) on different particle numbers $N_A$ at the final time of the calculation.
We computed the probability distribution to transfer multiple pairs of fermions during collisions happening with the coupling strength $v_0= 2\times10^{-3}g,\ 2\times10^{-2}g,\ 2\times10^{-1}g$ and $2g$ and plot the results in Fig.~\ref{fig:n1distrib_asym}.
We also show in Fig. \ref{fig:n1distrib_dE0p25} a situation where we clearly see interference patterns in the pair transfer probabilities. 
We see for all coupling strengths that the approach always gives the good order of magnitude for the addition or removal of 
one pair.  In particular, for 
strong coupling (Fig. \ref{fig:n1distrib_asym}-b), the \methodi captures very accurately the complete distribution including the 
interference effects seen in Fig. \ref{fig:n1distrib_dE0p25}. This nicely demonstrates the ability of the \methodi to grasp non-trivial 
interferences between the different mean-field trajectories that reproduces realistically the one observed in the exact solution.  
This is at variance with the semi-classical PSC method of Ref. \cite{Reg18} that was accurate only in the perturbative regime with the proper re-scaling of the 
interaction.  
In the perturbative regime ($v_0 \le 2\times10^{-2}g$), where the PSC method was quite accurate, we observe systematically that the \methodi  in general leads to overestimation 
of the small probabilities (multiple pair transfer in Fig. \ref{fig:n1distrib_asym}-a).  Note that this is also the case for the strong coupling regime for small probabilities that cannot be resolved in Fig. \ref{fig:n1distrib_asym}-b due to the linear scale used for the $y$-axis.

The discrepancy observed in the small coupling case and illustrated for instance in Fig. \ref{fig:n1distrib_asym} can have two origins. The first possible origin 
is numerical. To compute the \methodi evolution, we use a finite number of states $L$ as well as a lower cutoff on the eigenvalues obtained by diagonalizing the norm kernel. This cutoff determines the size of the active states (image of ${\cal N}$) used to invert ${\cal N}$. In practice the machine precision imposes a lower limit of this cutoff that prevents us to take into account components of the many body state associated with very small eigenvalues. These components could become of crucial importance when projecting the system into extreme configurations such as $N_A=0$ and this may explain the breaks in slope clearly visible for $v_0=2\times 10^{-3}$.

\begin{figure}[t!]
\includegraphics[width=0.95\linewidth]{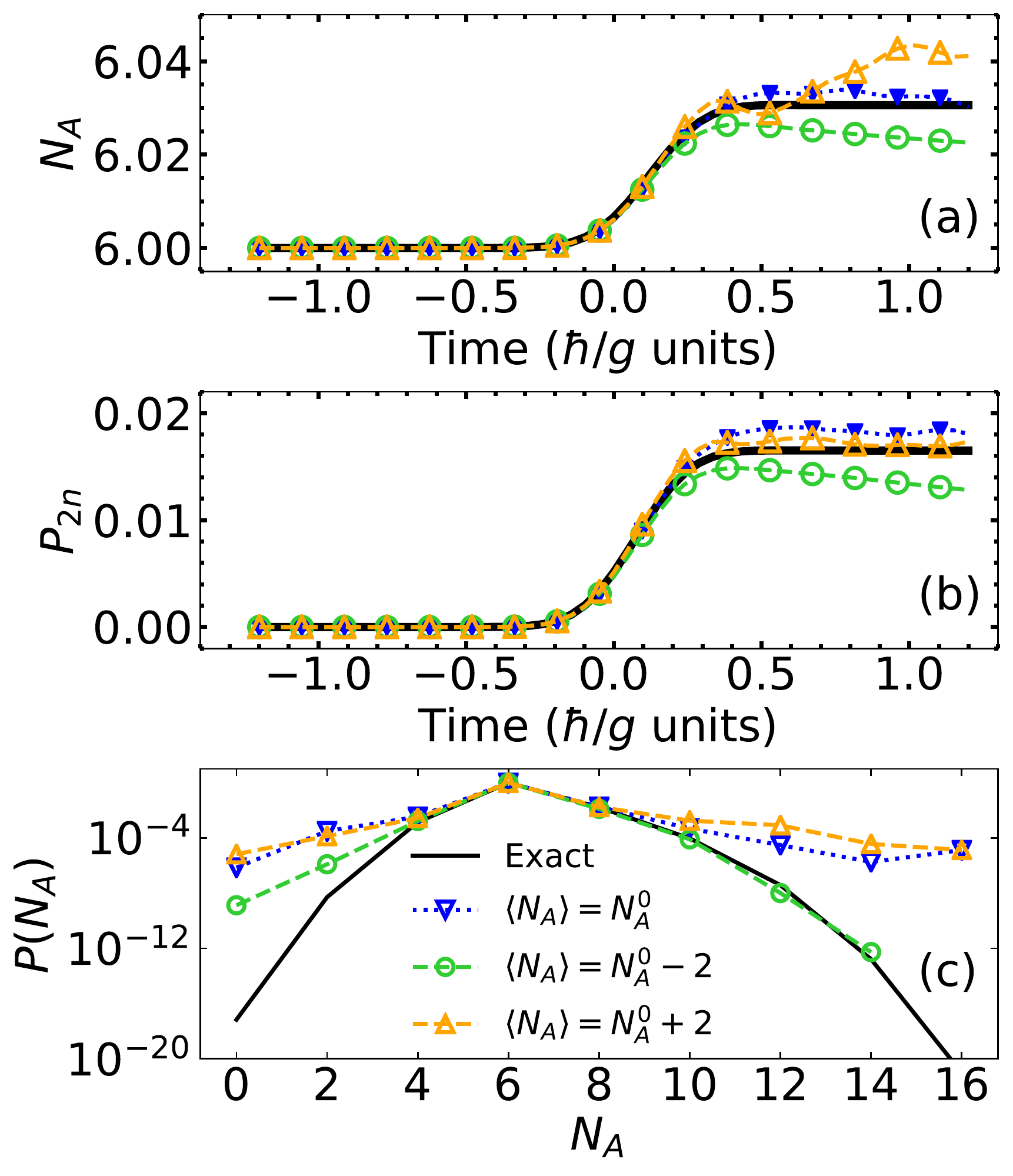}
 \caption{Illustration of the sensitivity of the average number of particles imposed initially in the subsystem $A$ (a), fluctuations $\sigma(N_A)$ (b) and extracted final pair transfer probabilities (c) 
 for the same physical situation as in Fig. \ref{fig:n1distrib_asym}  in the perturbative regime,  $v_0/g= 2\times 10^{-3}$.  
 The blue lower triangles correspond to the original case where $\langle N_A \rangle = N^0_A$ imposed in the initial vacuum to describe the sub-system $A$.  The green circles 
 and orange lower triangles correspond respectively to the initial constraints $\langle \hat N_A \rangle = N^0_A - 2$ and $N^0_A + 2$. In all cases, we imposed  $\langle \hat N_B \rangle. = N^0 - \langle \hat N_A \rangle$.  }
 \label{fig:allobs_asymnp}
\end{figure}

Beside the possible numerical issues, there is a source of error that is physical and is also inherent to the \methodi approach. The power of the method 
strongly depends on the possibility to describe a process in a very much reduced Hilbert space of states that account for the different channels relevant 
during the process under interest. 
In the present work, a quasi-particle state is used initially such that the mean particle number identifies with the physical
one, i.e. $\langle \hat N_A \rangle = N^0_A$ and $\langle \hat N_B \rangle = N^0_B$.
The variational space explored by \methodi is generated from this starting point.
Since the initial vaccum is optimized for the subspace $(N^0_A,N^0_b)$, we expect it to capture well the physics around this configuration. 
On the other hand, one might worry about the fact that it can properly generates the  physical states in other subspaces $(N_A, N_B)$ especially if $N_A$ and $N_B$ strongly differ from $N^0_A$  and $N^0_b$ as it is the case when several pairs are transferred.
The accuracy of the  \methodi predictions for multi-pair transfer directly reflects its ability to capture the physical output channels.

To test the sensitivity of pair transfer to many-body space explored by MC-TDHFB$_{\textrm{I}} $,  we performed several calculations in which we changed the average 
number of particles $\langle \hat N_A \rangle$ and $\langle \hat N_B \rangle$ imposed to the initial HFB ground states in each subsystem.
We kept here the average total number of particle fixed $\langle \hat N_A \rangle + \langle \hat N_B \rangle = N^0$. 
In each case, we still  consider the initial projection on $N^0$ and $N^0_A$ for the initial mixed state used in  MC-TDHFB$_{\textrm{I}}$. 
The effect of this procedure is to change the space of many body states probed by the \methodi variational principle. We clearly see in Fig. \ref{fig:allobs_asymnp} the sensitivity 
of the results, especially of the pair transfer probabilities, on the initial imposed particle number in $A$ and $B$. 
In some cases, we observe an improvement of the prediction of  small 
multi-pair transfer. 
However, we also see that alternative initial constraints usually deteriorate the prediction of the mean and fluctuations of the particle number $N_A$. 
Our conclusion is that, in the absence of a better criteria to constraint the initial state, the natural choice, i.e.  $\langle \hat N_A \rangle = N^0_A$ and $\langle \hat N_B \rangle = N^0_B$ remains the best choice.       

In summary, we have shown in the present section that the \methodi is predictive for the one pair transfer problem between small superfluid systems and this for a wide range of physical situations which include the ones that we encounter in deep sub-barrier collision of nuclear systems. It is also very accurate in the strong coupling regime 
and is able to describe quantum effects in the transfer probabilities. 
In the perturbative regime, it is accurate for the zero or one pair transfer process but leads to overestimation in the multi-pair transfer mechanism.   In this regime, we note 
however that it is possibility to cure this lacuna by using the combinatorial scheme as proposed in Ref.~\cite{Reg18} to extrapolate 
the probability distribution based on the \methodi predictions for $P_{-2n}, P_0$ and $P_{2n}$.
The fact that the method requires only on a few ($10$ to $30$ states) independent TDHFB trajectories to tackle properly a Hilbert space of dimension higher than $10^4$ sounds very appealing for realistic applications based on a Hamiltonian evolution.


\section{Conclusion}

In the present work, we discuss the possibility to use Multi-Configuration TDHFB approach to describe the transfer of particles between two finite superfluid systems. This approach has the advantage to incorporate the possible interference effects between different mean-field trajectories. It however also significantly increases the numerical effort.  We show here that, following the original idea of Ref. 
\cite{Rei83}, a possible simplification can be made by assuming that each TDHFB trajectory evolves independently from the others while 
treating the quantum mixture of these trajectories properly in time. This scheme is called MC-TDHFB$_{\textrm{I}} $. Specific formal and practical 
new aspects of the approach are discussed. The whole dynamics requires (i) the computation of a small ensemble of independent TDHFB trajectories (ii) the computation of a few observables kernel at each time step. 
The method is illustrated successfully for the problem of particle transfer between two superfluid systems.  
We tested this method for several cases of a simple model. In the case of a collision between two identical and degenerate system, our method provides the exact solution. In the more sophisticated case of two different and non-degenerated systems, the MC-TDHFB$_{\textrm{I}} $ solution  shows an excellent agreement of all the transfer observables as compared with the exact solution.
Comparing this method with phase-space combinatorial approach previously developed, we show that the quantum interferences between TDHFB trajectories explain some $40\%$ of the average drift of particles in a perturbative collision.

We would finally like to mention here that the direct application of \methodi to a realistic nuclear collision would still face a the difficulty that current implementations of the nuclear mean-field dynamics formalisms rely on an energy density functional and not a Hamiltonian. As reported in Ref.~\cite{Ben09,Rob10}, the non diagonal kernels are ill-defined in the context of multi-reference energy density functional and lead to divergences of the results. Although some prescriptions were proposed to circumvent this issue~\cite{Lac09} in the case of the particle number symmetry restoration, it is yet an open question how to soundly apply the general configuration mixing strategy with nuclear energy density functionals.

\section*{Acknowledgment}  
 We thank Y. Tanimura and  N. Hasegawa for pointing out the reference \cite{Rei83} before we started this work. We also thank G. 
 Scamps for regular discussions. This project has received funding from the European Union's Horizon 2020 research and innovation programme under Grant Agreement No. 654002.
\appendix

\section{Discussion on the derivation of the MC-TDHFB$_{\rm I}$ equation (\ref{eq:tdhw_proj})}
\label{sec:eom}

A standard difficulty in using an overcomplete set of states in the ansatz Eq.(\ref{eq:ansatz0}) is that there might exist several mixing functions $f(\bm{q}) $ that lead to the same state $|\psi\rangle$.
We must pay a special attention to this issue to perform properly the variational principle  on the action (\ref{eq:varia}). 
In this appendix, we discuss this topic, just after recalling some useful mathematical properties.

\subsection{Preliminary discussion: properties of the overlap matrix}
\label{sec:math_prop}

In the MC-TDHFB$_{\rm I}$ approach, at a given time, the mean-field states  $|\phi(\bm{q},t)\rangle$ are not orthogonal with each other.
As a consequence, there is no one-to-one mapping between the space $S(t)$ of many body states defined by the ansatz Eq.~(\ref{eq:ansatz0}) and the space $\mathcal{F}$ of the  functions of the label $\bm{q}$. For instance, any function $f$ such that
\begin{equation}
\forall q:\quad \left(\mathcal{N} f\right)(q) = \int_{q'} \mathcal{N}_{qq'} f(q') dq' = 0
\end{equation}
yields a many body state $|\psi\rangle =0$.
In the language of the linear algebra in the vectorial space $\mathcal{F}$, this means that any function that belongs to the kernel $\mathcal{K}(t)$ of the linear application $\mathcal{N}(t)$ gives the null many body states.
To handle this issue, it is useful to restrict the mixing functions $f$ to belong to the subspace $\mathcal{I}(t)$ (the image) of $\mathcal{N}$ which is orthogonal to $\mathcal{K}(t)$. At any time we have the decomposition
\begin{equation}
 \mathcal{F} = \mathcal{I}(t) \oplus \mathcal{K}(t).
\end{equation}
Although this splitting depends on time, we will drop the time dependency from the notations in what follows.
It is a standard results that there is a one-to-one mapping between the functions in $\mathcal{I}$ and the many body states in $S(t)$~\cite{Rin81}.
To derive the evolution equation for the mixing function, we introduce the projectors $\mathcal{P}$ and $\mathcal{Q}$ on the subspaces $\mathcal{I}$ and $\mathcal{K}$ respectively. The direct sum between these subspaces ensures
\begin{eqnarray}
{\cal P} + {\cal Q} & = & 1_{\cal F}.  \label{eq:comp}
\end{eqnarray} 
In practice, we determine these projectors by diagonalizing the overlap matrix $\mathcal{N}$. 
More precisely, let us assume that we diagonalize $\mathcal{N}$ leading to:
\begin{eqnarray}
{\cal N} &=& {\cal U} {\cal D} {\cal U}^{\dagger}, 
\end{eqnarray}  
with ${\cal D}$ diagonal. If the eigenvalues are sorted in ascending order, the first $d-r$ eigenvalues are zeros with $d$ the dimension of the matrices and $r$ the rank of $\mathcal{N}$. Then, the projectors explicitly read
\begin{align}
 \mathcal{Q}_{\bm{q}\bm{q'}} &= \sum_{i \le d-r} \mathcal{U}_{\bm{q}i}\mathcal{U}^\dagger_{i\bm{q}'} , \\
 \mathcal{P}_{\bm{q}\bm{q'}} &= \sum_{i > d-r} \mathcal{U}_{\bm{q}i}\mathcal{U}^\dagger_{i\bm{q}'}.
\end{align}
These projectors satisfy the identities:
\begin{equation}
 \mathcal{Q}\mathcal{N} =0, \quad \mathcal{P}\mathcal{N}=\mathcal{N}.
\end{equation}
Finally, it will also be useful to remark the properties
\begin{align}
\label{eq:ph_eq_h}
 \mathcal{P}\mathcal{H}=\mathcal{H},\quad \mathcal{P}\mathcal{H^{MF}} = \mathcal{H}^{MF}.
\end{align}
This actually holds for the kernel of any observable.

. 
 
\subsection{Derivation of Eq. (\ref{eq:tdhw_proj})}

Our aim is now to look for a mixing function $f(\bm{q})$ that makes the action  (\ref{eq:varia}) stationary.
\begin{equation}
 \delta S[f]=0.
\end{equation}
The variation $\delta S[f]$ should be performed on the set of function $f(\bm{q},t)$ that are (i) differentiable in time; (ii) leading to a state $|\psi(t)\rangle$ that is normalized at any time; (iii) belonging at any time to the subspace $\mathcal{I}(t)$ (the image of the overlap kernel as discussed in Sec.~\ref{sec:math_prop}). Using the projector $\mathcal{Q}$, this last condition formally reads ${\cal Q} f = 0$ or equivalently
\begin{align}
 ||\mathcal{Q}f||^2 = \int_{t \bm{q} \bm{q'}} f^*(\bm{q},t) \mathcal{Q}_{\bm{q}\bm{q}'}(t)f(\bm{q}',t) 
  =0. 
\end{align}
Here we used the compact notation $ \int_{t \bm{q} \bm{q'}} \cdots =  \int_{t \bm{q} \bm{q'}} \cdots dt d\bm{q} d\bm{q'}$.
In practice, we impose the last two constraints by minimizing the augmented action
\begin{align}
S'[f] &= \int_{t \bm{q} \bm{q'}}  f^*(\bm{q},t) 
\langle \phi(\bm{q},t)| H-i\hbar \partial_t |\phi(\bm{q}',t) \rangle f(\bm{q}',t) \nonumber \\
&+ \lambda_1 \int_t (||\psi(t)|| -1) dt \nonumber \\
&+ \lambda_2 \int_{t \bm{q} \bm{q'}} f^*(\bm{q},t) \mathcal{Q}_{\bm{q}\bm{q}'}(t)f(\bm{q}',t),
\end{align}
in the ensemble of functions differentiable in time. The $\lambda_1$ and $\lambda_2$ parameters are Lagrange multipliers that are determined by imposing the normalization and projection conditions respectively. This action is Hermitian, and by imposing $\partial S'[f]/\partial f^*(\bm{q},t)=0$, we find that the constrained variational principle is equivalent to the set of equations:
\begin{align}
f^\dagger \mathcal{N} f &= 1, \label{eq:norm} \\
\mathcal{Q} f &= 0 , \label{eq:qf} \\
i\hbar \mathcal{N} \dot{f} &= [\mathcal{H} - \mathcal{H}^{MF} + \lambda_1 \mathcal{N}] f  . \label{eq:hh}
\end{align}
%
To solve this equation in practice, we need the explicit expression of $\dot{f}$. The problem here is that $\mathcal{N}$ can only be inverted in the subspace ${\cal I}$ but $\dot{f}$ does not necessarily belongs to this subspace. 
To go further, we split $\dot{f}$ into its components in the two orthogonal subspaces $\mathcal{I}$ and $\mathcal{K}$:
\begin{equation}
 \dot{f}  = {\cal P} \dot{f} + {\cal Q} \dot{f}. 
\end{equation}
Inserting this decomposition, and projecting on both $\mathcal{K}$ and $\mathcal{I}$, we find that Eq. (\ref{eq:hh}) is equivalent to:
\begin{equation}
\left\{
\begin{array}{rl}
&    {\cal Q} [\mathcal{H} - \mathcal{H}^{MF} ] f = 0 \\
\\
&    i\hbar \mathcal{P}\dot{f} = \mathcal{N}^{-1}\mathcal{P}[\mathcal{H} - \mathcal{H}^{MF} + \lambda_1 \mathcal{N}]f
\end{array}
\right.
.
\end{equation}
With the property Eq.~(\ref{eq:ph_eq_h}), we can show that the first equation is already satisfied for any function $f$ and brings no constraint on the mixing function. In addition, the $\mathcal{P}$ projector on the right hand side of the second equation can safely be removed and we eventually obtain one equation for the $\mathcal{P}\dot{f}$ component.
The complementary contribution $ {\cal Q} \dot{f}$ to $ \dot{f}$ is immediately obtained using Eq. (\ref{eq:qf}) leading to:
\begin{eqnarray}
{\cal Q} \dot{f} &=& -\dot {\cal Q} {f} = +\dot {\cal P} f . \nonumber
\end{eqnarray}  
Summing up the two contributions, we deduce that $f$ must fulfill:
\begin{align}
i\hbar \dot{f} &= \left(\mathcal{N}^{-1} [\mathcal{H} - \mathcal{H}^{MF} ] + \lambda_1 \mathcal{P} \right) f 
+i\hbar \dot{\mathcal{P}} f  . \nonumber
\end{align}
Conversely, any function $f$ that satisfies the normalization and projection conditions at initial time and follows this evolution will verify Eq.~\cref{eq:norm,eq:qf,eq:hh} at any time. The stationarity of the augmented action is therefore equivalent to the first order equation:
\begin{align}
i\hbar \dot{f} &= \mathcal{N}^{-1} [\mathcal{H} - \mathcal{H}^{MF}] f  + \lambda_1 f
+i\hbar \dot{\mathcal{P}} f , 
\end{align}
with the initial conditions
\begin{align}
 f^\dagger(t_0) \mathcal{N}(t_0)f(t_0) &=1, \\
 \mathcal{Q}(t_0)f(t_0) = 0.
\end{align}
Note that since the $\lambda_1$ term introduces just a phase, it could actually be removed without affecting the dynamics which leads to Eq. (\ref{eq:tdhw_proj}).


\end{document}